\documentstyle[aps,prl,preprint,floats,epsfig]{revtex}

\newcommand{\unitsP}  {GeV/$c$}
\newcommand{\unitsM}  {GeV}

\begin{document}

\preprint{\tighten\vbox{\hbox{\hfil CLNS 01/1752}
                        \hbox{\hfil CLEO 01-17}
}}

\title{Hadronic Mass Moments in Inclusive Semileptonic $B$ Meson
Decays}  

\author{CLEO Collaboration}
\date{\today}

\maketitle
\tighten

\begin{abstract} 
We have measured the first and second moments of the hadronic
mass-squared distribution in $B \rightarrow X_c \ell \nu $,
for $ P_{lepton} > 1.5$ \unitsP.  We find
$\langle M_X^2 - \bar M_D^2 \rangle = 0.251 \pm 0.066\ {\rm GeV}^2$,
$\langle (M_X^2 -  \langle M_X^2 \rangle)^2 \rangle = 
0.576 \pm 0.170\ {\rm GeV}^4$,
where
$\bar M_D$ is the spin-averaged $D$ meson mass.  From that first moment and
the first moment of the photon energy spectrum in $b \rightarrow s \gamma$,
we find the HQET parameter $\lambda_1$
($\overline {MS}$, to order $1/M_B^3$ and $\beta_0 \alpha_s^2$)
to be $ -0.24 \pm 0.11\ {\rm GeV}^2$.  Using these first moments and the $B$
semileptonic width, and  assuming parton-hadron duality,
we obtain $\vert V_{cb}\vert = 0.0404 \pm 0.0013$.
\end{abstract}
\newpage

{
\renewcommand{\thefootnote}{\fnsymbol{footnote}}

\begin{center}
D.~Cronin-Hennessy,$^{1}$ A.L.~Lyon,$^{1}$ S.~Roberts,$^{1}$
E.~H.~Thorndike,$^{1}$
T.~E.~Coan,$^{2}$ V.~Fadeyev,$^{2}$ Y.~S.~Gao,$^{2}$
Y.~Maravin,$^{2}$ I.~Narsky,$^{2}$ R.~Stroynowski,$^{2}$
% ...
\end{center}
 
\small
\begin{center}
$^{1}${University of Rochester, Rochester, New York 14627}\\
$^{2}${Southern Methodist University, Dallas, Texas 75275}\\
% ...
\end{center}

\setcounter{footnote}{0}
}
\newpage

    The heavy quark limit of QCD\cite{theory} is potentially a very
useful tool for relating measured inclusive properties in $B$ meson decay,
such as semileptonic branching fractions, to fundamental CKM parameters like
$V_{cb}$ and $V_{ub}$.  The expressions for inclusive observables are expansions
in inverse powers of the $B$ meson mass $M_B$~\cite{Adam,Gremm,Voloshin}.  At
order $1/M_B$, the non-perturbative parameter
$\overline{\Lambda}$ enters, and at order $1/M_B^2$ two more parameters,
$\lambda_{1}$
and $\lambda_{2}$ appear.  Intuitively, these parameters may be thought of
as the
energy of the light quark and gluon degrees of freedom ($\overline{\Lambda}$),
the  average momentum-squared of the $b$ quark (--$\lambda_{1}$), 
and the energy of the
hyperfine interaction of the spin of the $b$ quark with the light degrees of
freedom ($\lambda_{2}/M_B$).  The parameter $\lambda_{2}$ can be extracted
directly from the $B^* - B$ mass splitting\cite{Gremm}.  The
other two parameters can be obtained from inclusive measurement or
calculated theoretically with techniques capable of handling non-perturbative
effects, such as lattice QCD\cite{Kronfeld}.

    There are two problems associated with the interpretation of
measured inclusive properties, one associated with the convergence of the
expansion, and another with the validity of the assumptions underlying the
expansion.  The inclusive observables are
expansions in powers of $1/M_B$, and at each order more non-perturbative
parameters appear.  By order $1/M_B^2$ there are three parameters and at
order $1/M_B^3$ another six parameters, $\rho_1$, $\rho_2$,
${\cal T}_1$ -- ${\cal T}_4$.  Without good
estimates for the additional parameters we must rely on the rapid convergence
of the expansion.
    The other problem is the validity of the assumption of
parton-hadron duality implicit in this approach, and its potential for
introducing additional uncertainties not included in the present
estimates\cite{Isgur}.  Thus, the experimental determination of
$\overline{\Lambda}$
and $\lambda_{1}$ with several different methods is necessary to support the
validity of parton-hadron duality\cite{Bigi}.

    Much interest has been raised by the possibility of estimating
$\overline{\Lambda}$ and $\lambda_{1}$ using hadronic spectral moments in
semileptonic $B$ decays\cite{Adam,Gremm,Voloshin}.
In this Letter we report a measurement of the first and second moments of the
distribution in the hadronic mass-squared in the inclusive
semileptonic decay $b \rightarrow c \ell \nu $.
For this analysis, the leptons are restricted to the kinematical region 
$P_{\ell}$ $\ge$ 1.5 \unitsP. In particular, we report
measurements of $\langle M_X^2 - \bar M_D^2 \rangle $ and
$\langle (M_X^2 - \bar M_D^2)^2 \rangle $, where $M^2_X$ is the mass-squared of
the charmed hadronic system $X_c$, and $\bar M_D$ is the spin-averaged $D$ meson
mass, $0.25M_D + 0.75 M_{D^*} = 1.975$ \unitsM.  The theoretical expansion for
these two observables has been carried out to order $1/M_B^3$ and order
$\beta_0 \alpha_s^2$ in the $\overline {MS}$ renormalization
scheme~\cite{Adam,Gremm}.
(Here $\beta_0 = (33 - 2 n_f)/3 = 25/3$ is the one-loop QCD beta function.)
We also report the second moment taken about the first moment
rather than about $\bar M_D^2$, i.e.,
$\langle (M_X^2 -  \langle M_X^2 \rangle)^2 \rangle $, the mean square width of
the mass-squared distribution.
(The theoretical expansion for this is readily obtained from those for
$\langle M_X^2 - \bar M_D^2 \rangle $ and
$\langle (M_X^2 - \bar M_D^2)^2 \rangle $.)  We use the first moment, along with
the first moment of the photon energy spectrum in
$b \rightarrow s \gamma$\cite{preceeding}, to
obtain $\lambda_1$ and an improved extraction of $V_{cb}$ from the $B$ meson
semileptonic width.


The data used in this analysis were taken with the CLEO
detector\cite{CLEO-detector} at the Cornell Electron Storage Ring (CESR), and
consist of 3.2 ${\rm fb}^{-1}$ at the $\Upsilon(4S)$ resonance and
1.6 ${\rm fb}^{-1}$  at a center-of-mass energy 60 MeV below the resonance.
The sample contains 3.4 million $B \bar B$ pairs.  We select events containing
a lepton -- $\mu$ or $e$ -- with momentum between 1.5 and 2.5 \unitsP.  We
``reconstruct'' the neutrino in the event using energy and momentum
conservation
of the entire event, exploiting the hermiticity of the CLEO detector.
The neutrino energy is taken as the difference of twice the beam energy
and the sum of the energies of all detected particles,
while the neutrino momentum is the
negative of the vector sum of the momenta of all detected particles.
Considerable effort was expended to remove double counting between 
calorimeter and tracking chamber measurements.   
To assure a well-measured neutrino, we require: a
neutrino mass consistent with zero; no additional leptons in the event;
a measured net charge of zero for the event.  The ``neutrino reconstruction''
aspect of this analysis is similar to that of Ref.~\cite{pi-ell-nu}, and is 
described in detail in Ref.~\cite{SER}. Event shape requirements
are applied to distinguish  the jetty event environment typical of 
$e^+ e^- \rightarrow q \overline{q} $
light quark pair production from the more isotropic environment of
$e^+ e^- \rightarrow B \overline{B} $
events. We achieve a sample consisting of 
89\%  $e^+ e^- \rightarrow B \overline{B} $ and 11\% from the
continuum, with an efficiency for the desired events of $\approx$2\%.
The desired semileptonic $B$ decays, $b \rightarrow c \ell \nu $,
represent 95\%  of the $e^+ e^- \rightarrow B \overline{B} $ sample while the
remaining  consists of (2.8 $\pm$ 0.6)\% secondary lepton production (from Monte
Carlo simulation) and (2.1 $\pm$ 1.1)\% $b \rightarrow u \ell \nu $ (using
$\vert V_{ub}/V_{cb}\vert = 0.07 \pm 0.02$).

We determine the mass of the hadronic system $X$ in
$B \rightarrow X_c \ell \nu$
from the lepton and neutrino momentum vectors alone:
\begin{eqnarray}
M_X^2 &=& (E_B - E_\ell - E_\nu)^2 - (\vec P_B - \vec P_\ell - \vec P_\nu)^2
\nonumber \\
 &=& M_B^2 + M_{\ell \nu}^2 - 2 E_B E_{\ell \nu} + 2 \vert \vec P_B
\vert \vert \vec P_{\ell \nu} \vert \cos \theta_{\ell \nu, B}\ .
\label{eq:xmass}
\end{eqnarray}

\noindent For $B$ mesons produced at the $\Upsilon (4S)$, $E_B$ and
$\vert \vec P_B \vert$ are known and constant, but the angle between the $B$ and
$\ell \nu$ system varies
from event to event, and is not known.  Since $\vert \vec P_B \vert$ is small
(300 MeV/c), we approximate $M_X^2$ by dropping the final term in
Eq.~\ref{eq:xmass}, writing
\begin{equation}
\widetilde {M_X^2} = M_B^2 + M_{\ell \nu}^2 - 2 E_B E_{\ell \nu}.
\end{equation}

\begin{figure}
\hspace*{1.5cm}
\epsfxsize = 3.25in
\epsffile{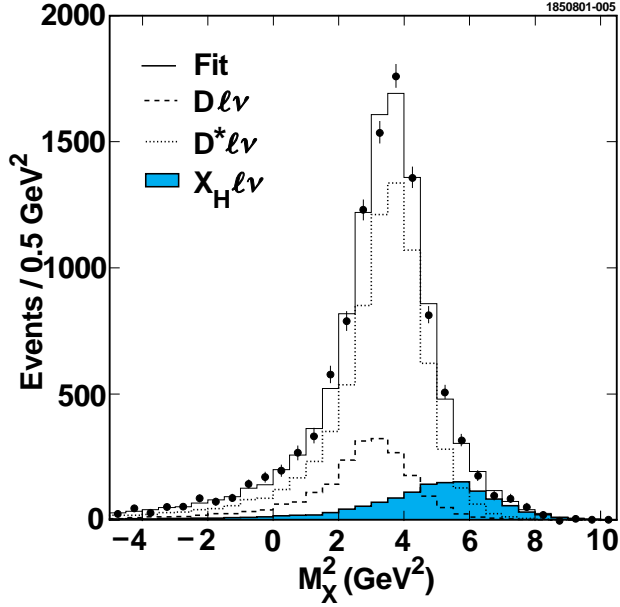}
\caption{\ Measured $ \widetilde {M_X^2}$ distributions, for 
background-corrected data (points), Monte Carlo (solid line),
and the three components of the Monte Carlo: $B \rightarrow D \ell \nu $
(dashed), $B \rightarrow D^* \ell \nu $ (dotted), 
$B \rightarrow X_H \ell \nu $ (shaded). The normalization of each component
is derived from a fit  to the data.
\label{fig:mxave}}
\end{figure}

        The background-subtracted $\widetilde {M_X^2}$ distribution,
consisting of 11900 $B$ meson decays, is shown in Fig.~\ref{fig:mxave}. 
The background from  continuum
events has been subtracted using the data collected below the  $\Upsilon(4S)$
resonance, scaled to the luminosity of the on-resonance data and corrected
for the dependence of the production cross section on beam energy.  The
small backgrounds from secondary lepton sources and from
$b \rightarrow u \ell \nu $ decays, which we obtain from Monte Carlo
simulation, have also been subtracted.

For the purpose of extracting the 
moments of the ${M_X^2}$ distribution,
we divide the $b \rightarrow c \ell \nu $ decays into three components:
$B \rightarrow D \ell \nu $, $B \rightarrow D^* \ell \nu $, and
$B \rightarrow X_H \ell \nu $ where
$X_H$ represents all the high-mass charmed meson resonances as well as the 
charmed non-resonant decays.
The individual components are shown in Fig.~\ref{fig:mxave}.
We use measured form factors~\cite{CLEO-D*-ff}
to model the  $B \rightarrow D \ell \nu $ and $B \rightarrow D^* \ell \nu $
 decays.
The true ${M_X^2}$ distributions for $B \rightarrow D \ell \nu $
 and $B \rightarrow D^* \ell \nu $ are narrow resonances
at $M_D^2$ and $M_{D^*}^2$. 
The widths of the Monte Carlo predictions in Fig.~\ref{fig:mxave} 
for these resonances are dominated by neutrino energy-momentum resolution and
our neglect of the last term in Eq.~\ref{eq:xmass}.
The high-mass contribution, $B \rightarrow X_H \ell \nu $, 
is modeled using six resonances above the $D^{*}$ with the decay 
properties specified by ISGW2 form factors~\cite{ISGW2}, and also non-resonant
multi-body final states such as $B \rightarrow D \pi \ell \nu $ and 
$B \rightarrow D^* \pi \ell \nu $, which are decayed according to the
prescription of  Goity and Roberts~\cite{GR}.

A fit of the  Monte Carlo to the data $\widetilde {M_X^2}$
distribution determines the relative 
contributions from $B \rightarrow D \ell \nu $,
$B \rightarrow D^* \ell \nu $ and $B \rightarrow X_H \ell \nu $.
The relative rates  
and the generated masses  are used to calculate
$\langle M_X^2 - \bar M_D^2 \rangle $ and 
$\langle (M_X^2 - \bar M_D^2)^2 \rangle $ of the true ${M_X^2}$ distribution.
Equation~\ref{eq:sh} shows the derivation of the average mass squared,
$M_X^2$, from the relative rates. 

 \begin{equation}
\label{eq:sh}
\langle M_X^2 \rangle = r_D \cdot M_D^2 + r_{D^*} \cdot M_{D^*}^2 + r_{X_H}
\cdot \langle M_{X_H}^2 \rangle,
\end{equation}

\noindent where $r_D$ is the rate of $B \rightarrow D \ell \nu $
production compared to 
the combined rate of $B \rightarrow D \ell \nu $, $B \rightarrow D^* \ell \nu $,
$B \rightarrow X_H \ell \nu $, and similarly for $r_{D^*}$ and $r_{X_H}$.  The
individual values obtained for $r_D$, $r_{D^*}$ and $r_{X_H}$, while perfectly
consistent with world average branching fractions\cite{PDG}, are not well
determined and
are sensitive to the model chosen for $B \rightarrow X_H \ell \nu$.  The
{\it moments}, however, are well-determined and stable against model changes, as
discussed below.
We find
$\langle M_X^2 - \bar M_D^2 \rangle \equiv M1 =
0.251 \pm 0.023 \pm 0.062\ {\rm GeV}^2$,
$\langle (M_X^2 - \bar M_D^2)^2 \rangle \equiv M2 = 0.639 \pm 0.056 \pm 0.178
\ {\rm GeV}^4$, and
$\langle (M_X^2 -  \langle M_X^2 \rangle)^2 \rangle \equiv M2^\prime =
0.576 \pm 0.048 \pm 0.163\ {\rm GeV}^4$, where the errors are statistical and
systematic, in that order.  The experimental errors on 
$\langle (M_X^2 -  \langle M_X^2 \rangle)^2 \rangle $ are somewhat smaller
than for $\langle (M_X^2 - \bar M_D^2)^2 \rangle$ and have a smaller correlation
with the first moment.  (A correction for final state radiation, not included in
the Monte Carlo samples used in our fits, has been applied, using
PHOTOS\cite{Photos}.)

The errors on both first and second moments are  dominated by systematic errors.
The leading contribution is from the simulation parameters that impact neutrino
resolution: photon identification efficiency, tracking efficiency, and the rate
of additional neutrals such as $K^0_L$ and additional neutrinos; it amounts to
$\pm$0.058 ${\rm GeV}^2$, $\pm$0.140 ${\rm GeV}^4$, and
$\pm$0.129 ${\rm GeV}^4$, for $M1$, $M2$, and $M2^\prime$, respectively.

The second leading source of systematic error is from the models for the 
high-mass contribution to the $\widetilde {M_X^2}$ distribution.  
We have varied 
aspects of the high-mass component in order to quantify the sensitivity. The
six contributing mass states of the resonant component
(above $D^{*}$) have been systematically dropped singly, in
pairs and in triplets so as to vary the internal structure of the resonant
model.  Taking the r.m.s. deviations of these variations, we find errors of
$\pm$0.015 GeV$^2$,  $\pm$0.090 GeV$^4$, and $\pm$0.083 GeV$^4$, for $M1$, $M2$,
$M2^\prime$.

Another contributing uncertainty arises from the 
lack of knowledge on the amount and shape of non-resonant contribution 
to the high-mass component. Although we fix the fraction of non-resonant
to resonant high-mass states during a fit, we systematically
vary this fraction over the limits that the data allow. A one unit 
variation of fit $\chi^2$ determines a systematic variation of 
0.011 GeV$^2$ for $M1$ and 0.060 GeV$^4$ and 0.054 GeV$^4$ 
for $M2$ and $M2^\prime$.  Systematic errors other than those from neutrino
resolution simulation, modelling of high-mass resonances, and modelling of non-resonant
high-mass decays, such as the subtractions for secondary leptons and for
$b \rightarrow u \ell \nu$, the final state radiation correction, and the
$B \rightarrow D^{(*)} \ell \nu$ form factor uncertainties, are negligible by
comparison.

As an alternative to the default Goity-Roberts parameterization, 
we have also used a phase space model to generate the four-body non-resonant 
decays~\cite{SER}. This phase space model generates, on average,
higher mass states than the Goity-Roberts parameterization but yields 
hadronic mass moments consistent with those obtained from the Goity-Roberts
parameterization. This 
observation emphasizes the fact that the data essentially constrain
the product of the average mass squared and production rate while
these quantities may individually vary significantly. 

    The correlation coefficients between errors of first and second moments are
positive, and substantial.  They are +0.71 for $M1$ -- $M2$ (+0.56 for $M1$ --
$M2^\prime$) for the statistical error,
+0.50 (+0.34) for the systematic error, and +0.52 (+0.36) for the total error.

  The expressions \cite{Adam,Falketal} for the hadronic mass moments in
$B \rightarrow X_c \ell \nu$, to order $\beta_0 \alpha^2_s$ and $1/M_B^3$,
subject to the restriction $P_\ell > 1.5$ \unitsP, are given in
Eqs.~\ref{eq:hadmom1} and \ref{eq:hadmom3}. (Due to technical difficulties, the
coefficients of the $\frac{\bar \Lambda}{M_B} \frac{\alpha_s}{\pi}$ terms were
computed {\it without} the 1.5 GeV lepton energy restriction, and so are only
approximate, believed \cite{Falketal} good to $\pm$ 50\%.)  
\begin{eqnarray}
    \frac{\langle M_X^2 - \bar M_D^2 \rangle}{M_B^2} = & 
[0.0272 \frac{\alpha_s}{\pi}
+ 0.058 \beta_0 \frac{\alpha_s^2}{\pi^2}
+ 0.207 \frac{\bar \Lambda}{\bar M_B}(1+0.43 \frac{\alpha_s}{\pi}) +
0.193 \frac{\bar \Lambda^2}{\bar M_B^2} 
+ 1.38 \frac{\lambda_1}{\bar M_B^2} + 0.203
\frac{\lambda_2}{\bar M_B^2} \nonumber \\
& +0.19 \frac{\bar \Lambda^3}{\bar M_B^3}
+ 3.2 \frac{\bar \Lambda \lambda_1}{\bar M_B^3}
+1.4 \frac{\bar \Lambda \lambda_2}{\bar M_B^3} \nonumber \\
& +4.3 \frac{\rho_1}{\bar M_B^3}  -0.56 \frac{\rho_2}{\bar M_B^3}
+ 2.0 \frac {{\cal T}_1}{\bar M_B^3}
+ 1.8 \frac {{\cal T}_2}{\bar M_B^3} + 1.7 \frac {{\cal T}_3}{\bar M_B^3}
+ 0.91 \frac {{\cal T}_4}{\bar M_B^3} + {\cal O}(1/\bar M^4_B)],
\label{eq:hadmom1}
\end{eqnarray}

\begin{eqnarray}
\frac{\langle (M_X^2 - \langle M_X^2 \rangle )^2 \rangle}{M_B^4} = &
[0.00148 \frac{\alpha_s}{\pi} + 0.0025\beta_0 \frac{\alpha_s^2}{\pi^2}
+ 0.027 \frac{\bar \Lambda}{\bar M_B} \frac{\alpha_s}{\pi} + 0.0107 \frac{\bar
\Lambda^2}{\bar M_B^2} -0.12 \frac{\lambda_1}{\bar M_B^2} \nonumber \\
& +0.02 \frac{\bar \Lambda^3}{\bar M_B^3}
- 0.06 \frac{\bar \Lambda \lambda_1}{\bar M_B^3}
-0.129 \frac{\bar \Lambda \lambda_2}{\bar M_B^3} \nonumber \\
& -1.2 \frac{\rho_1}{\bar M_B^3}  +0.0032 \frac{\rho_2}{\bar M_B^3} 
- 0.12 \frac {{\cal T}_1}{\bar M_B^3} - 0.36 \frac {{\cal T}_2}{\bar M_B^3}
+ {\cal O}(1/\bar M^4_B)],
\label{eq:hadmom3}
\end{eqnarray}

\noindent In these expressions, $\bar M_B$ represents the spin-averaged $B$
meson mass, 5.313 \unitsM.  

\begin{figure}[h]
\hspace*{1.5cm}
\epsfxsize = 3.25in
\epsffile{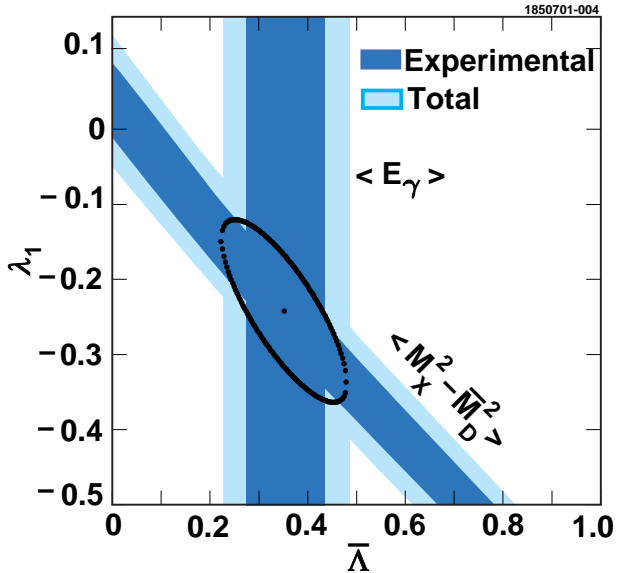}
\caption{Bands in $\bar \Lambda$ -- $\lambda_1$ space defined by 
$\langle M_X^2 - \bar M_D^2 \rangle$ , the measured first moment of hadronic
mass-squared, and $\langle E_\gamma \rangle$, the first moment of the photon
energy spectrum in $b \rightarrow s \gamma$
\protect \cite{preceeding}.
The inner bands  
indicate the error bands from the measurements.  The light grey extensions
include the errors from theory. All bands are derived from 
$\cal{O}$$(1/\bar M_B^3)$
$\cal{O}$$(\beta_0 \alpha_s^2)$
HQET expressions, using the $\overline {MS}$ renormalization scheme.
\label{fig:bands}}
\end{figure}

    The $1/M_B^3$ parameters $\rho_i$, ${\cal T}_i$ are
estimated\cite{Gremm}, from dimensional considerations,  to be
$\sim (0.5 {\rm GeV})^3$.  Taking values of $\rho_2$ and ${\cal T}_1$ through
${\cal T}_4$ to be $0.0 \pm (0.5 {\rm GeV})^3$, taking $\rho_1$ (believed to be
positive) to be
$\frac{1}{2}(0.5 {\rm GeV})^3 \pm \frac{1}{2}(0.5 {\rm GeV})^3$, taking
$\lambda_2 = 0.128\ \pm\ 0.010\ {\rm GeV}^2$ (appropriate with a calculation to
order $1/M_B^3$)\cite{Gremm}, and using $\alpha_s(m_b)$ = 0.220, the expressions combined with our measurements define bands in
$\bar \Lambda - \lambda_1$ space.  The band for the first moment is shown in
Fig.~\ref{fig:bands}.
The dark grey region indicates the error band from the measurement; the light
grey extension includes the error from the theoretical expression, in particular
from the $\rho_1 - {\cal T}_4$ terms and from the scale uncertainty
($\alpha_s(m_b/2)$ = 0.275 to $\alpha_s(2 m_b)$ = 0.176).

    In the preceeding Letter\cite{preceeding}, we presented measurements of the
first and second moments of the photon energy spectrum in
$b \rightarrow s \gamma$, and gave the OPE expansion expressions for those
moments, again valid to order $\beta_0 \alpha^2_s$ and $1/M_B^3$.  Again,
equation plus measurement defines a band in $\bar \Lambda - \lambda_1$ space.
The band for the first
moment, $\langle E_\gamma \rangle$ is also shown in Fig.~\ref{fig:bands}.
    The expressions for the second moments converge more slowly in $1/M_B$
than those for the first moments,  and the theoretical advice \cite{Falketal}
is {\it not} to put much trust in the bands they define.  Consequently we have
not shown them in Fig.~\ref{fig:bands}.

The intersection of the two bands from the first moments determines
$\bar \Lambda$ and $\lambda_1$.  A $\Delta \chi^2 = 1$ ellipse is shown.  The
values obtained are

$$\bar \Lambda = 0.35 \pm 0.07 \pm 0.10\ {\rm GeV}\ ,$$

$$ \lambda_1 = -0.236 \pm 0.071 \pm 0.078\ {\rm GeV}^2\ .$$

\noindent Here, the first error is from the experimental error on the
determination of the two moments, and the second error from the theoretical
expressions.  (Using the information from all four bands, first and second
moments, the results differ little, both as to central values and as to
errors.)  Note that $\bar \Lambda$ and $\lambda_1$ are scheme and order
dependent.  The values obtained above are for $\bar \Lambda$ and $\lambda_1$
to order $1/M^3$, order $\beta_0 \alpha_s^2$, in the $\overline {MS}$
renormalization scheme.

    Given this determination of $\bar \Lambda$ and $\lambda_1$, we can use them
to improve the determination of $\vert V_{cb} \vert$ from the measured
$B \rightarrow X_c \ell \nu$ semileptonic width.  The
expression \cite{semi-theory,Gremm} for the semileptonic width,  to order
$\beta_0 \alpha^2_s$ and $1/M_B^3$, is given in Eq.~\ref{eq:vcb}.
\begin{eqnarray}
\Gamma_{sl} = & \frac{G^2_F \vert V_{cb} \vert^2 M_B^5}{192 \pi^3}0.3689
[1 - 1.54 \frac{\alpha_s}{\pi} - 1.43 \beta_0 \frac{\alpha_s^2}{\pi^2}
 - 1.648 \frac{\bar \Lambda}{M_B}(1 -
0.87 \frac{\alpha_s}{\pi} ) - 0.946\frac{\bar \Lambda^2}{M_B^2} 
 -3.185 \frac{\lambda_1}{M_B^2}  \nonumber \\
&+ 0.02 \frac{\lambda_2}{M_B^2} 
-0.298 \frac{\bar \Lambda^3}{M_B^3} - 3.28 \frac{\bar \Lambda \lambda_1}{M_B^3}
+10.47 \frac{\bar \Lambda \lambda_2}{M_B^3} 
 -6.153 \frac{\rho_1}{M_B^3}  +7.482 \frac{\rho_2}{M_B^3} \nonumber \\
&- 7.4 \frac {{\cal T}_1}{M_B^3} + 1.491 \frac {{\cal T}_2}{M_B^3}
-10.41 \frac {{\cal T}_3}{M_B^3}
-7.482 \frac {{\cal T}_4}{M_B^3}
+ {\cal O}(1/M^4_B)]\ .
\label{eq:vcb}
\end{eqnarray}

    For the experimental determination of $\Gamma_{sl}$, we use:
${\cal B}(B \rightarrow X_c \ell \nu) = (10.39 \pm 0.46)\% $ \cite{semi},
$\tau_{B^\pm} = (1.548 \pm 0.032)$ ps \cite{PDG},
$\tau_{B^0} = (1.653 \pm 0.028)$ ps \cite{PDG},
$f_{+-}/f_{00} = 1.04 \pm 0.08$ \cite{Sylvia},
giving $\Gamma_{sl} = (0.427 \pm 0.020) \times 10^{-10}$ MeV.

    Combining the measured semileptonic width with the theoretical expression
for it, and using the determination of $\bar \Lambda$ and $\lambda_1$ from the
first moments, we find

$$\vert V_{cb} \vert = (4.04 \pm 0.09 \pm 0.05 \pm 0.08) \times 10^{-2} \ ,$$

\noindent where the errors are from experimental determination of $\Gamma_{sl}$,
from experimental determination of $\bar \Lambda$ and $\lambda_1$, and from the
$1/M_B^3$ terms and scale uncertainty in $\alpha_s$, in that order.  This
gives a determination of
$\vert V_{cb} \vert $ from inclusive processes, with a precision of $\pm$3.2\%.
This result depends on the assumption of global parton-hadron duality, with
its unknown uncertainties.

    Summarizing, we have measured the first and second moments 
of the hadronic mass-squared distribution in the $B$ meson semileptonic 
decay to charm, $B \rightarrow X_c \ell \nu $.  We find
$\langle M_X^2 - \bar M_D^2 \rangle  = 0.251 \pm 0.023 \pm 0.062\ {\rm GeV}^2$, 
$\langle (M_X^2 - \bar M_D^2)^2 \rangle =
0.639 \pm 0.056 \pm 0.178\ {\rm GeV}^4$,
and $\langle (M_X^2 -  \langle M_X^2 \rangle)^2 \rangle =
0.576 \pm 0.048 \pm 0.163\ {\rm GeV}^4$.
The measurement of $\langle M_X^2 - \bar M_D^2 \rangle $
and the HQET expression for this moment are used, in conjunction with similar
information on the first moment of the photon energy spectrum in
$b \rightarrow s \gamma$, to determine $\lambda_{1}$ and $\overline{\Lambda}$.
These in turn are used, along with the $B$ meson semileptonic width,
to obtain $V_{cb}$.


We gratefully acknowledge the effort of the CESR staff in providing us 
with excellent luminosity and running conditions.
We thank Adam Falk and Michael Luke for performing calculations for us.
We thank Falk, Luke, M. Gremm and A. Kapustin, and M. Voloshin,
A.I. Vainshtein, and M. Shifman and Z. Ligeti for informative discussions
and correspondences.
This work was supported by
the National Science Foundation,
the U.S. Dept. of Energy, 
Research Corporation,
and the Texas Advanced Research Program.

\end{document}